\documentclass[preprint,aps,pra,amsmath,amssymb,floatfix,showpacs]{revtex4-1}
\usepackage{graphicx}
\usepackage{dcolumn}
\usepackage{bm}

\newcommand{\nt}{Nature}
\newcommand{\lp}{Laser Phys.}
\newcommand{\lpl}{Laser Phys. Lett.}
\begin{document}

\title{Bistable behavior of a two-mode Bose-Einstein condensate in an optical cavity}

\author{S. Safaei$^{1}$}
\email{safaei@fen.bilkent.edu.tr}
\author{\"O. E. M\"ustecapl{\i}o\u{g}lu$^{2}$}
\author{B. Tanatar$^{1}$}
\affiliation{$^{1}$Department of Physics, Bilkent University, 06800 Ankara, Turkey\\
$^{2}$Department of Physics, Ko\c{c} University, Sar{\i}yer, 34450 Istanbul , Turkey}
\begin{abstract}
We consider a two-component Bose-Einstein condensate in a one-dimensional optical cavity. 
Specifically, the condensate atoms are taken to be in two degenerate modes due to their 
internal hyperfine spin degrees of freedom and they are coupled to the cavity field and 
an external transverse laser field in a Raman scheme. A parallel laser is also exciting 
the cavity mode. When the pump laser is far detuned from its resonance atomic transition 
frequency, an effective nonlinear optical model of the cavity-condensate system is developed 
under Discrete Mode Approximation (DMA), while matter-field coupling has been considered 
beyond the Rotating Wave Approximation. By analytical and numerical solutions of the 
nonlinear dynamical equations, we examine the mean cavity field and population difference 
(magnetization) of the condensate modes. The stationary solutions of both the mean cavity 
field and normalized magnetization demonstrate bistable behavior under certain conditions 
for the laser pump intensity and matter-field coupling strength.
\\\\
Topic: Physics of Cold Trapped Atoms (Report number: 6.6.4)
\end{abstract}
\maketitle
\section{Introduction}
\label{intro}

Recently, due to experimental advances in coupling a dilute gas of bosons 
to a single mode of an optical cavity, many theoretical and experimental works 
have been performed in order to explore and explain the physics of such complex 
systems. Bose-Einstein condensate (BEC) itself is a very rich platform which allows us 
to examine different properties of quantum systems such as quantum turbulence~
\cite{turbulence1,turbulence2,turbulence3}, quantum chaos~\cite{chaos} and 
entanglement~\cite{entanglement}. A Bose-Einstein condensate can also be used to 
study nonequilibrium dynamics and decoherence in finite quantum systems~\cite{equilibration} 
as well as calculation of fluctuation indices for atomic systems~\cite{indices}. 
Moreover, if confined in an optical lattice~\cite{yukalov09}, Bose-Einstein 
condensates provide the possibility to observe solid-state physics processes, 
such as localization~\cite{localization1,localization2}, in these atomic systems. 
Further uses of BECs are exemplified in transport problems~\cite{ad_transport},
interaction driven instabilities~\cite{critical_point}, stability of boson-fermion 
gaseous mixtures~\cite{bfmix}, and Raman pump-probe experiments~\cite{raman_pump_probe}.
 
In a more complex setup, where a dilute condensate of bosons is confined inside 
a high-finesse optical cavity subject to external laser fields, if the cavity 
mode and laser fields are far-detuned from the atomic transition frequency of 
the condensate atoms, system is in dispersive regime. Under the dispersive regime 
conditions, atom-field interaction provides an optical lattice for the condensate 
atoms which affects their mechanical motion. On the other hand, the atoms cause a 
position-dependent phase-shift of the cavity mode. As a result, the condensate-cavity 
system is highly nonlinear with nonlocal nonlinearities which give rise to a series 
of interesting physical phenomena such as self-organization of condensate atoms, 
cavity enhanced superradiant scattering, Dicke quantum phase transition~\cite{nagy-prl10,
esslinger-nature10} and optical bistability. The nonlinearity caused by atom-filed 
interaction can be more effective~\cite{zubairy11} than the nonlinearity from atom-atom 
interaction, which itself can play crucial role in scattering dynamics of 
condensates~\cite{reflection}.

The optical bistability, which is the focus of this article, has been studied 
in spinor BECs \cite{zhou09,zhou10} and two-component BEC with two modes coupled 
by a classical field \cite{dong11}. In a system consisting of single-mode 
BEC in an optical cavity, a transverse laser pump has also been used to control 
the bistability of cavity photons induced by a parallel pump \cite{zubairy11}. 
In this work we consider a two-mode BEC in a one-dimensional optical cavity, 
where two laser fields are applied to the system, one parallel to the cavity 
axis and the other one perpendicular to it. Specifically we assume that the 
transverse pump is scattered to the cavity mode by condensate atoms in the Raman 
scheme \cite{cola04, uys-meystre07}. Under this condition, we examine the effect 
of transverse pump strength on the bistability of both mean cavity photon number 
and normalized population difference (magnetization) of the two modes. Stationary 
state of cavity field and condensate wavefunction are obtained under Discrete 
Mode Approximation (DMA) which has been shown to be reliable for similar systems 
\cite{zubairy11,zhang09}.

This paper is organized as follows. In Sec.\ref{model} first the model for our system is 
introduced, then the Hamiltonian of the system and equations of motion of cavity field 
and condensate are derived. Using DMA, we solve the equations of motion for the steady 
state of the system in Sec.\ref{results}, where we show how mutual bistability of the mean 
cavity photon number and magnetization take place under certain conditions for laser field 
intensities and cavity-atom coupling strength. Finally, we summarize our work in 
Sec.\ref{summary}.

\section{Two-mode BEC in one dimensional cavity}
\label{model}

We consider a condensate of $N$ atoms, each with two internal degrees of freedom, 
shown as states $b$ and $c$ and an excited state $e$, in a one dimensional cavity
along the $x$ axis. The cavity has a single mode with frequency $\omega_c$ and is 
subjected to laser field with frequency $\omega_0$, which is far detuned from the 
atomic transition, in directions parallel and perpendicular to its axis. If the 
transverse laser field interacts with matter through Raman scattering (Fig.~1), 
then the full Hamiltonian of the system will have following form 

\begin{eqnarray}
\label{eq:H1}
H&=&\sum_{j=b,c}\int dx~\psi^{\dag}_j\left(-\frac{\hbar^2}{2m}\frac{\partial^2}{\partial x^2}
+ V_j(x) + \hbar\omega_{bc}\delta_{j,c}\right)\psi_j
+ \sum_{i,j=b,c} \int dx~\frac{u_{ij}}{2}\psi^{\dag}_i\psi_i\psi^{\dag}_j\psi_j\nonumber\\
&+&\hbar\omega_c a^{\dag}a-i\hbar\eta_{||}(ae^{i\omega_0t}-a^{\dag}e^{-i\omega_0t})+H_{Raman},
\end{eqnarray}
where $\omega_{bc}$ is the frequency of transition between modes $b$ and $c$ and {$u_{ij}$} 
are the interaction strengths of atoms in modes $i$ and $j$. The parallel laser field 
intensity is shown by $\eta_{||}$ and $a$ and $a^{\dag}$ are the annihilation and creation 
operators of the cavity mode. The interaction of atoms with the transverse pump is shown by 
the Raman scattering Hamiltonian ($H_{Raman}$) and has the following form

\begin{eqnarray}
\label{eq:Hraman1}
H_{Raman}=-i\hbar\int dx~\psi^{\dag}_e h_0
(e^{-i\omega_0t}+e^{i\omega_0t})
\psi_b + H.c.\nonumber\\
-i\hbar\int dx~\psi^{\dag}_e g_0\cos(kx)(a+a^{\dag})\psi_c + H.c.,
\end{eqnarray}
where $h_0$ and $g_0$ are the atom-pump and atom-cavity coupling strengths, respectively, and 
the Hamiltonian is written without Rotating Wave Approximation. After adiabatically eliminating 
the excited state $\psi_e$ and introducing $U_0=g_0^2/\Delta_0$ and $\eta=h_0g_0/\Delta_0$, with 
$\Delta_0=\omega_0-\omega_{be}$ being the pump detuning from the atomic transition, one obtains

\begin{eqnarray}
\label{eq:Hraman2}
H_{Raman}&=&\hbar\eta \int dx~(a+a^{\dag})(e^{-i\omega_0 t}+e^{i\omega_0 t})
\cos(kx)\left(\psi^{\dag}_c \psi_b + \psi^{\dag}_b \psi_c\right)\nonumber\\
&+&\frac{\hbar h_0^2}{\Delta_0}\int dx~(e^{-i\omega_0 t}+e^{i\omega_0 t})^2\psi^{\dag}_b \psi_b
\nonumber\\&+&\hbar U_0\int dx~\cos^2(kx)(a+a^{\dag})^2 \psi^{\dag}_c\psi_c.
\end{eqnarray}
\begin{figure}[t]
\centering
\includegraphics[width=0.8\textwidth]{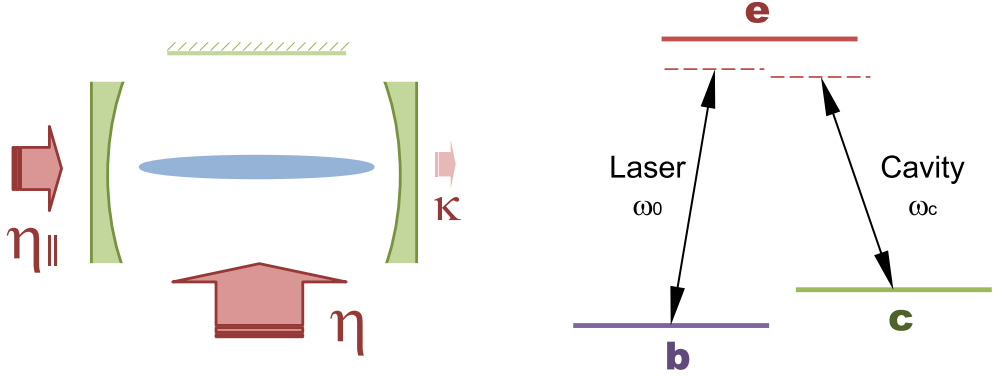}
\label{fig:raman}
\caption{Left: Schematic drawing of a BEC in one dimensional optical cavity subject to 
parallel and transverse laser fields. Cavity has a decay rate of $\kappa$. Right: Two 
internal modes ($b$ and $c$) of the BEC atoms are coupled by cavity field and laser 
field via the atomic excited state $e$ in Raman scattering manner. Both laser field and 
cavity field are detuned from the atomic transition frequency.}
\end{figure}
Substituting (\ref{eq:Hraman2}) into (\ref{eq:H1}), moving to a rotating frame defined by 
the unitary operator $U=e^{-i\omega_0 t a^{\dag}a}$,
and ignoring two-photon processes result in

\begin{eqnarray}
\label{eq:H2}
H = \sum_{j=b,c}\int dx~\psi^{\dag}_j{\cal{H}}\psi_j 
+ \sum_{i,j=b,c} \int dx~\frac{u_{ij}}{2}\psi^{\dag}_i\psi_i\psi^{\dag}_j\psi_j,
\end{eqnarray}
for the Hamiltonian of the system where

\begin{eqnarray}
\label{eq:Hsingle}
{\cal{H}}&=&-\frac{\hbar^2}{2m}\frac{\partial^2}{\partial x^2} 
+\left(\hbar U_0\cos^2(kx)(aa^{\dag}+a^{\dag}a)+V_c(x)+\hbar\omega_{bc}\right)\sigma^+\sigma^- 
\nonumber\\&+&\left(\frac{2\hbar h_0^2}{\Delta_0}+V_b(x)\right)\sigma^-\sigma^+ 
+\hbar\eta(a+a^{\dag})\cos(kx)(\sigma^- + \sigma^+)\nonumber\\
&-&\hbar\delta_c a^{\dag}a-i\hbar\eta_{||}(a-a^{\dag}).
\end{eqnarray}
Here $\sigma^+=|\psi_c\times\psi_b|$ and $\sigma^-=|\psi_b\times\psi_c|$ are the ascending 
and descending operators in the two-mode manifold and $\delta_c=\omega_0-\omega_c$ is the 
pump-cavity detuning. From equations (\ref{eq:H2}) and (\ref{eq:Hsingle}) one can derive the 
Heisenberg equations of motion for condensate modes and cavity field as follow

\begin{eqnarray}
\label{eq:heisen-psib}
\dot{\psi}_b=
&-&\frac{i}{\hbar}\left(
-\frac{\hbar^2}{2m}\frac{\partial^2}{\partial x^2}+V_b(x)
+\frac{2\hbar h_0^2}{\Delta_0}
+u_{bb}\psi^{\dag}_b\psi_b + u_{bc}\psi^{\dag}_c\psi_c
\right)\psi_b\nonumber\\
&-&i\eta\cos(kx)(a+a^{\dag})\psi_c,
\end{eqnarray}

\begin{eqnarray}
\label{eq:heisen-psic}
\dot{\psi}_c=
&-&\frac{i}{\hbar}\left(
-\frac{\hbar^2}{2m}\frac{\partial^2}{\partial x^2}+V_c(x)
+\hbar\omega_{bc}+\frac{\hbar g_0^2}{\Delta_0}\cos^2(kx)(aa^{\dag}+a^{\dag}a)\right)\psi_c
\nonumber\\
&-&\frac{i}{\hbar}\left(u_{cc}\psi^{\dag}_c\psi_c + u_{bc}\psi^{\dag}_b\psi_b\right)\psi_c
-i\eta\cos(kx)(a+a^{\dag})\psi_b,
\end{eqnarray}
and

\begin{eqnarray}
\label{eq:heisen-alpha}
\dot{a}&=&
i\left(i\kappa+\delta_c-\frac{2g_0^2}{\Delta_0}\int dx~\psi^{\dag}_c\cos^2(kx)\psi_c\right)a
\nonumber\\&-&i\eta\int dx~\cos(kx)(\psi^{\dag}_c \psi_b + \psi^{\dag}_b \psi_c)+\eta_{||},
\end{eqnarray}
where we have introduced a decay rate $\kappa$ for the cavity in (\ref{eq:heisen-alpha}).
In order to obtain the equations of motion for the expectation values of field operators, 
we further simplify our system. Since we work with a condensate of large number of atoms, 
we can safely treat the condensate as a coherent state and disentangle it from photons. 
We also neglect higher order correlations within the photon subsystem and use $<a><a^{\dag}>$ 
and $<a^{\dag}><a>$ instead of $<aa^{\dag}>$ and $<a^{\dag}a>$. At the end, as far as we 
only deal with the intensities, we treat the photon as a coherent state as well. As a result, 
following equations of motion are obtained

\begin{eqnarray}
\label{eq:eom-psib}
\dot{\psi}_b=
&-&\frac{i}{\hbar}\left(
-\frac{\hbar^2}{2m}\frac{\partial^2}{\partial x^2} + V_b(x)
+\frac{2\hbar h_0^2}{\Delta_0}
+u_{bb}|\psi_b|^2 + u_{bc}|\psi_c|^2
\right)\psi_b\nonumber\\
&-&2i\eta\cos(kx)\alpha_r\psi_c,
\end{eqnarray}

\begin{eqnarray}
\label{eq:eom-psic}
\dot{\psi}_c=
&-&\frac{i}{\hbar}\left(
-\frac{\hbar^2}{2m}\frac{\partial^2}{\partial x^2} + V_c(x) + \hbar\omega_{bc}
+2\hbar U_0\cos^2(kx)|\alpha|^2\right)\psi_c\nonumber\\
&-&\frac{i}{\hbar}\left(u_{cc}|\psi_c|^2 + u_{bc}|\psi_b|^2\right)\psi_c
-2i\eta\cos(kx)\alpha_r\psi_b,
\end{eqnarray}

\begin{eqnarray}
\label{eq:eom-alpha}
\dot{\alpha}&=&
i\left(i\kappa+\delta_c-2U_0\int dx~|\psi_c|^2\cos^2(kx)\right)\alpha
\nonumber\\&-&i\eta\int dx~\cos(kx)(\psi^*_c \psi_b + \psi^*_b\psi_c)+\eta_{||},
\end{eqnarray}
where $\alpha_r$ is the real part of the cavity field $\alpha$.

\section{Bistability of photon number and magnetization}
\label{results}

To solve the equations of motion derived in Sec.\ref{model} for the steady state of the system, 
we employ Discrete Mode Approximation. The ground state of the condensate, without any laser 
field, is a homogeneous macroscopic state with zero momentum, which we refer to by $\phi_0$. 
Above superradiance threshold one can assume that the condensate is fragmented to a symmetric 
superposition of the states with $\pm\hbar k$ momentum due to the transverse laser field. On 
the other hand, since the cavity mode is excited by the parallel pump, absorption and emission 
of the cavity photons can excite the condensate to the superposition  of states with $\pm 2\hbar k$ 
momentum. Therefore if we consider first order perturbation on the homogeneous wavefunction 
$\phi_0$, following functions can be used as the basis for DMA:

\begin{eqnarray}
\phi_0=\sqrt{1/L}\nonumber\\
\phi_1=\sqrt{2/L}\cos(kx)\nonumber\\
\phi_2=\sqrt{2/L}\cos(2kx)
\end{eqnarray}
with $L$ being the length of the cavity (condensate). Now the wavefunctions of the 
two modes of the condensate can be expanded in this basis as follow

\begin{eqnarray}
\psi_b(x,t)=\sum_{i=0}^2{\phi_i b_i}\nonumber\\
\psi_c(x,t)=\sum_{i=0}^2{\phi_i c_i},
\end{eqnarray}

If we substitute these wavefunctions into the equations of motion of the two modes, 
by ignoring external potentials and atom-atom interaction, the equation of motion 
for the condensate wavefunction can be written in a compact form:
 
\begin{eqnarray}
i\hbar\frac{d}{dt}X=H(\alpha)X=[H_0+H_1+H_2+2\alpha H_3+2\alpha_rH_4]X,
\end{eqnarray}
where $X=(b_0,b_1,b_2,c_0,c_1,c_2)^T$ and

\begin{eqnarray}
H_0=\hbar\omega_r
\left(\begin{matrix}
0&0&0&0&0&0\\
0&1&0&0&0&0\\
0&0&4&0&0&0\\
0&0&0&0&0&0\\
0&0&0&0&1&0\\
0&0&0&0&0&4
\end{matrix}\right)
\end{eqnarray}

\begin{eqnarray}
H_1=\frac{2\hbar h_0^2}{\Delta_0}
\left(\begin{matrix}
1&0&0&0&0&0\\
0&1&0&0&0&0\\
0&0&1&0&0&0\\
0&0&0&0&0&0\\
0&0&0&0&0&0\\
0&0&0&0&0&0
\end{matrix}\right)
\end{eqnarray}

\begin{eqnarray}
H_2=\hbar\omega_{bc}
\left(\begin{matrix}
0&0&0&0&0&0\\
0&0&0&0&0&0\\
0&0&0&0&0&0\\
0&0&0&1&0&0\\
0&0&0&0&1&0\\
0&0&0&0&0&1
\end{matrix}\right)
\end{eqnarray}

\begin{eqnarray}
H_3=\frac{\hbar U_0}{4}
\left(\begin{matrix}
0&0&0&0&0&0\\
0&0&0&0&0&0\\
0&0&0&0&0&0\\
0&0&0&2&0&\sqrt{2}\\
0&0&0&0&3&0\\
0&0&0&\sqrt{2}&0&2
\end{matrix}\right)
\end{eqnarray}

\begin{eqnarray}
H_4=\frac{\hbar\eta}{2}
\left(\begin{matrix}
0&0&0&0&\sqrt{2}&0\\
0&0&0&\sqrt{2}&0&1\\
0&0&0&0&1&0\\
0&\sqrt{2}&0&0&0&0\\
\sqrt{2}&0&1&0&0&0\\
0&1&0&0&0&0
\end{matrix}\right).
\end{eqnarray}
Here $\omega_r=\hbar k^2/2m$ is the recoil frequency.

To examine the equilibrium properties of the system, we set $\dot{\alpha}=0$ 
in (\ref{eq:eom-alpha}) which results in

\begin{eqnarray}
\label{eq:steadyalpha1}
\alpha=\frac{\eta\int dx~\cos(kx)(\psi_c^*\psi_b+\psi_b^*\psi_c)+i\eta_{||}}
{i\kappa+\delta_c-2U_0\int dx~\cos^2(kx)\psi_c^*\psi_c}.
\end{eqnarray}
On the other hand, one can easily obtain the following alternative expressions for 
the integrals in the above equation

\begin{eqnarray}
\int dx~\cos^2(kx)\psi_c^*\psi_c=\frac{1}{\hbar U_0}X^{\dag}H_3X
\end{eqnarray}

\begin{eqnarray}
\int dx~\cos(kx)(\psi_b^*\psi_c+\psi_c^*\psi_b)=\frac{1}{\hbar\eta}X^{\dag}H_4X.
\end{eqnarray}
Therefore cavity field $\alpha$ and averaged photon number $n=|\alpha|^2$ will read as

\begin{eqnarray}
\label{eq:steadyalpha2}
\alpha=\frac{\frac{1}{\hbar}X^{\dag}H_4X+i\eta_{||}}{i\kappa+\delta_c -\frac{2}{\hbar}X^{\dag}H_3X}
\end{eqnarray} 

\begin{eqnarray}
\label{eq:steadyn}
n=\frac{(\frac{1}{\hbar}X^{\dag}H_4X)^2+\eta_{||}^2}{\kappa^2+(\delta_c -\frac{2}{\hbar}X^{\dag}H_3X)^2}.
\end{eqnarray}

\begin{figure}[t]
\centering
\includegraphics[width=1.0\textwidth]{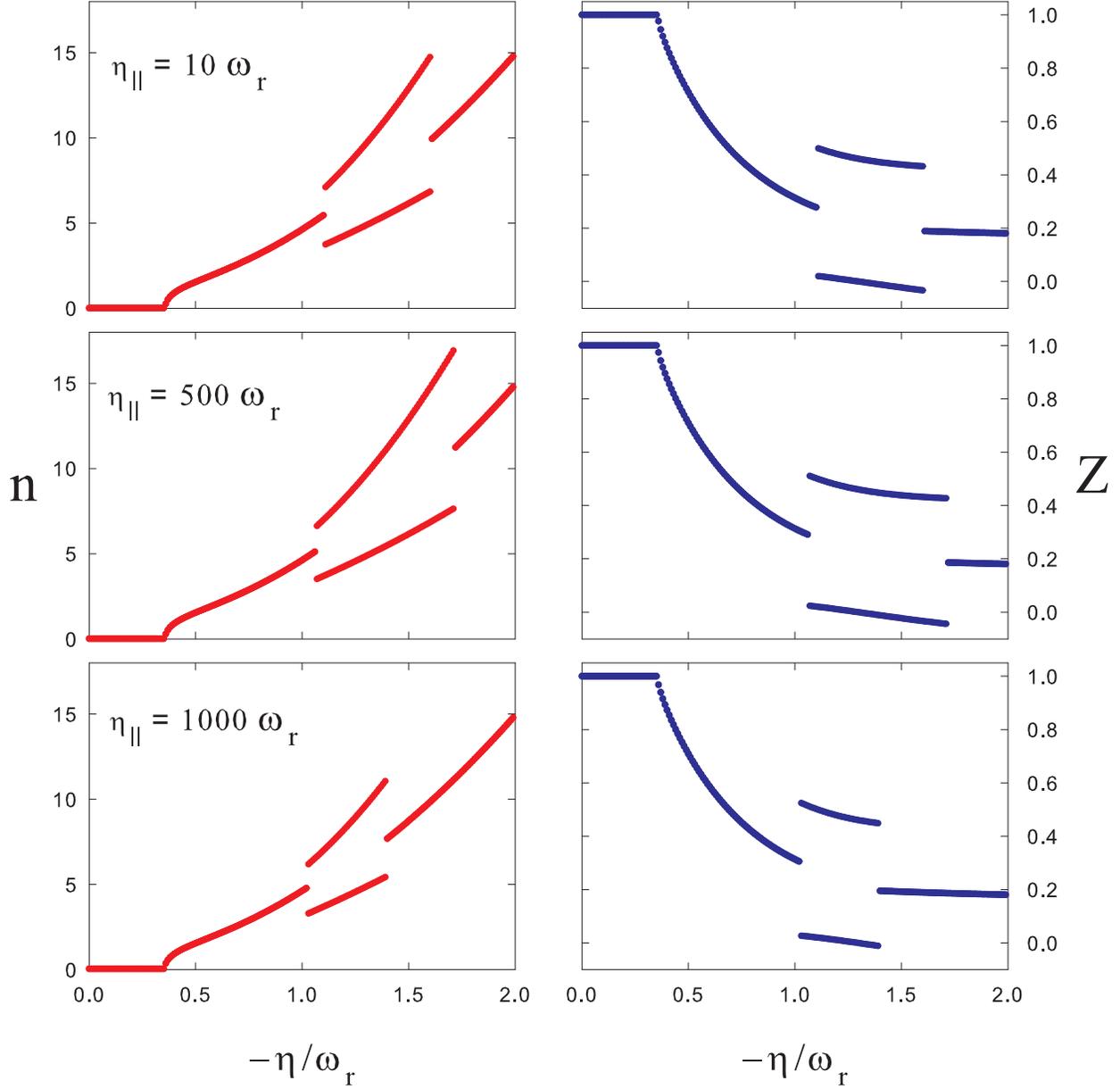}
\caption{Mean cavity field $n$ (left) and normalized magnetization $Z$ (right) as functions 
of $\eta$ for different values of parallel pump strength $\eta_{||}$. We remind that $\eta$ 
is proportional to transverse pump strength and atom-cavity coupling strength. In all plots 
$U_0=-0.5~\omega_r$, $\Delta_0=-4\times 10^6~\omega_r$, $\kappa=400~\omega_r$, 
$\delta_c=4800~\omega_r$, $N=4.8\times 10^4$, and $\omega_{bc}=\omega_r$.}
\end{figure}

\begin{figure}[t]
\centering
\includegraphics[width=1.0\textwidth]{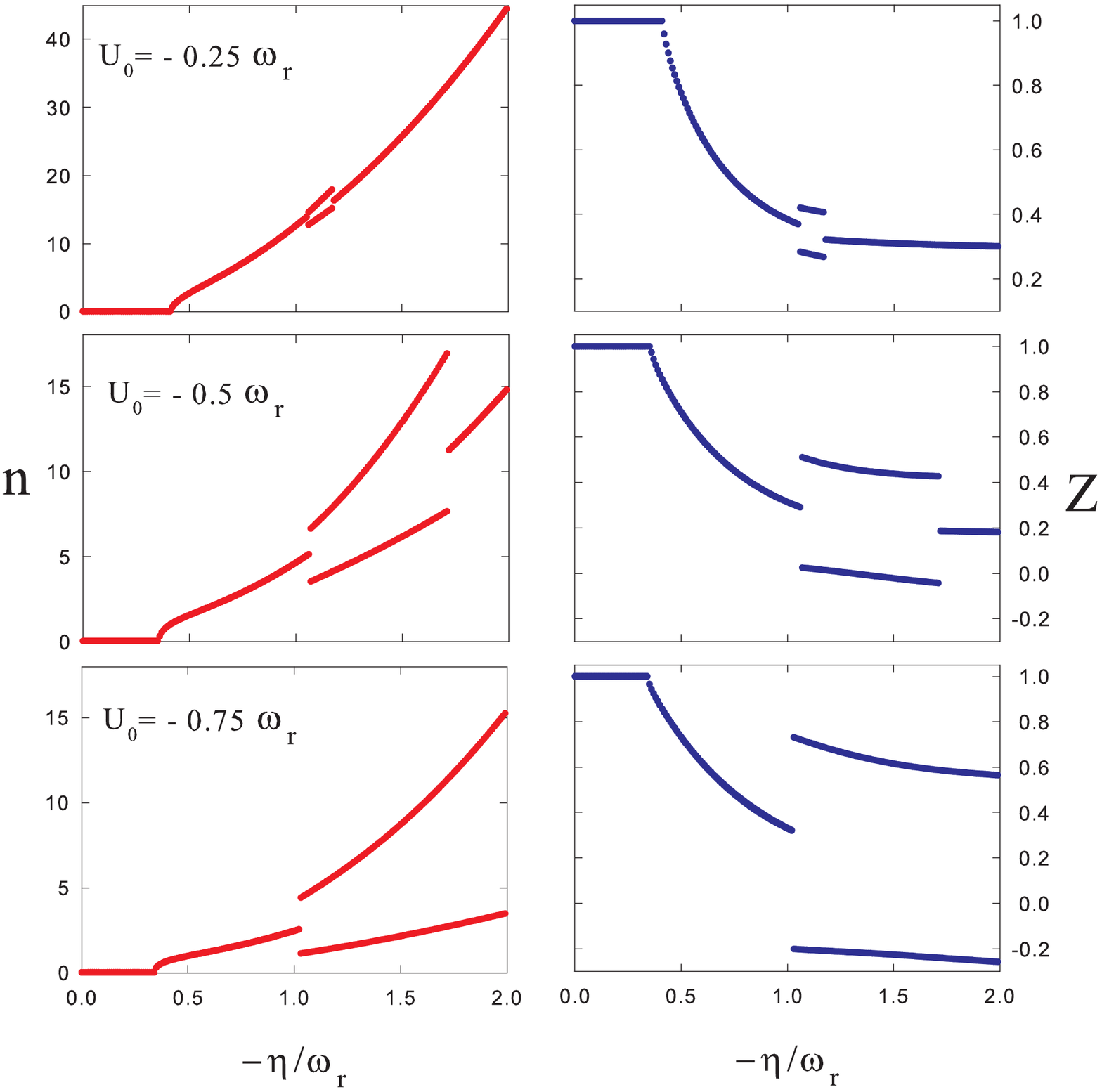}
\caption{Mean cavity field $n$ (left) and normalized magnetization $Z$ (right) as functions 
of $\eta$ for different values of atom-cavity coupling strength $U_0$. In all plots 
$\eta_{||}=500~\omega_r$, $\Delta_0=-4\times 10^6~\omega_r$, $\kappa=400~\omega_r$, 
$\delta_c=4800~\omega_r$, $N=4.8\times 10^4$, and $\omega_{bc}=\omega_r$.}
\end{figure}

Here one should notice that if there is no transverse laser field ($h_0=0$) then $\eta=0$ and 
$H_4=0$. As a result, there will be no transition between the two modes and if atoms are 
initially in the ground state $b$, they will stay there. Therefore expectation values of both 
$H_4$ and $H_3$ in (\ref{eq:steadyn}) or, equivalently, the integrals in (\ref{eq:steadyalpha1}) 
would be zero and no bistability in the system is expected. The importance of the transverse 
laser pump could be predicted from the crucial role it plays in Raman scattering Hamiltonian. 
On the other hand to better understand the contribution of parallel laser field in bistability, 
we remind that due to dependence of $X$ on $\alpha$, expectation values in numerator and 
denominator of (\ref{eq:steadyn}) are functions of $n$. To lowest order, we can assume that 
these terms depend on $n$ linearly. Therefore (\ref{eq:steadyn}) is a cubic equation of $n$ 
which in the case of $\eta_{||}=0$ has always a zero root. Therefore, to avoid a zero root in 
(\ref{eq:steadyn}), we will consider cases where $\eta_{||}\neq0$.

To obtain the mean cavity field $n$ for the steady state of the system we need to 
find the wavefunction $X_s$ in steady state. Since $H(\alpha)X_s=E_0X_s$ is nonlinear 
due to the dependence of $X$ on $\alpha$, we first solve it for $E_0$ and $X_s$ with 
a guess for the value of $\alpha$. Then by substituting the resulting $X_s$ into 
(\ref{eq:steadyalpha2}), a new $\alpha$ is obtained. If this new $\alpha$ is equal to 
the guessed value then the steady state is reached, otherwise we repeat the procedure, 
while using this new $\alpha$ as the guess, until the steady state is attained.

In addition to the average photon number, in a system consisting of two-mode condensate, 
there is another quantity which reveals the nonlinear effects of matter-field interaction 
and that is the normalized population difference (magnetization $Z$) of two modes. In our 
system magnetization is defined as $Z=\int dx(|\psi_b|^2-|\psi_c|^2)/N$, where 
$N=\int dx(|\psi_b|^2+|\psi_c|^2)$ is the total number of atoms and is fixed.

Fig.~2 shows mean cavity field $n$ and normalized magnetization $Z$ as functions 
of $\eta$ for different values of parallel pump strength $\eta_{||}$. We have considered a 
condensate of $N=48000$ atoms in a cavity with decay rate $\kappa=400~\omega_r$. The laser 
field is detuned from cavity mode by $\delta_c=4800~\omega_r$ and from atomic transition by
$\Delta_0=-4\times 10^6~\omega_r$. In all plots, atom-cavity coupling is assumed to be 
$U_0=-0.5~\omega_r$. As one can observe from these plots, the effect of parallel pump on 
widening the area of bistability is not monotonic. While increasing $\eta_{||}$ from $10$ to 
$500$ results in a wider interval of $\eta$ in which bistability happens, further increasing 
it to $1000$ has the opposite effect. Moreover, $\eta_{||}$ has larger impact on the width 
of the area in which bistability takes place than on the values of $n$ and $Z$ in bistable 
region. The interesting point about this system is the fact that nonlinear effects of 
matter-field coupling result in bistable behavior of both matter and field, such that, 
bistability happens for magnetization of condensate atoms exactly at the same region of $\eta$ 
where $n$ has shown bistable behavior. Moreover, it can be seen in Fig.~2 that, in some points, 
two states not only with different values of magnetization but also with different signs of 
magnetization are stable.

In the next step, to better understand the role of atom-cavity coupling on bistable behavior 
of the system, we keep the value of parallel pump strength $\eta_{||}$ constant and increase 
the strength of atom-cavity coupling. Fig.~3 shows the average photon number $n$ and normalized 
magnetization $Z$ as functions of $\eta$ for the same values of parameters used in Fig.~2 but 
this time with $\eta_{||}=500~\omega_r$ and for three different values of $U_0$. Clearly, 
atom-cavity coupling has a significant effect on the width of the area in which bistability 
happens. More importantly, $U_0$ can change the distance between the two stable branches. It is 
worth mentioning that $\eta$ itself is proportional to $g_0$ and therefore for a fixed value of 
$\eta$, larger $U_0$ means smaller value for $h_0$. As a result, increasing $U_0$ makes it possible 
to achieve bistability with smaller values and wider range of atom-laser coupling $h_0$.

\section{Conclusion \label{summary}}
In this work we developed an effective Hamiltonian and equations of motion for a 
cavity-condensate system consisting of a two-mode BEC in a one dimensional cavity, while 
parallel and transverse laser fields are applied to the system. Under DMA, simultaneous 
and mutual bistability (multistability) of cavity field and population difference 
(magnetization) of the two modes has been observed for different values of transverse 
and parallel pump strength. The system shows bistable behavior for quite wide range of 
parameters. Moreover it has been shown that, with strong enough cavity-matter coupling 
strength, bistability happens between two states with different sign of magnetization.

\acknowledgments
The authors would like to thank Y. Castin and M. \"O. Oktel for fruitful discussions and 
S. Safaei would like to acknowledge discussions with P. Elahi. This work was supported by 
T\"UB\.{I}TAK (Grant. No. 109T267).


\end{document}